\documentclass[a4paper,11pt]{article}
\pdfoutput=1 % if your are submitting a pdflatex (i.e. if you have
\usepackage{jheppub_modified} % for details on the use of the package, please
\usepackage[T1]{fontenc} % if needed

\usepackage{verbatim}
\usepackage{color}
\usepackage{graphicx}
\usepackage{subcaption}

\usepackage{bigcenter}

\usepackage{sectsty}
\allsectionsfont{\boldmath}

\usepackage{xspace}
\usepackage[dvipsnames,table]{xcolor}
\usepackage{tabularx}
\newcolumntype{C}[1]{>{\centering\arraybackslash}p{#1}}

\newcommand{\eq}[1]{Eq.~(\ref{#1})}
\newcommand{\bib}[1]{Ref.~\cite{#1}}
\newcommand{\refs}[1]{Refs.~\cite{#1}}
\newcommand{\bibs}[1]{\cite{#1}}
\newcommand{\fig}[1]{Fig.~\ref{#1}}
\newcommand{\tab}[1]{Table~\ref{#1}}

\newcommand{\sect}[1]{Section~\ref{#1}}

\newcommand{\bea}{\begin{eqnarray}}
\newcommand{\eea}{\end{eqnarray}}

\newcommand{\nn}{\nonumber}
\newcommand{\crn}{\nonumber \\}

\newcommand{\fr}{\frac}

\newcommand{\gev}{{\unskip\,\text{GeV}}}

\title{Enhancing the doubly-longitudinal polarization 
in $WZ$ production at the LHC}

\author{Thi Nhung Dao,}
\author{Duc Ninh Le}

\affiliation{Faculty of Fundamental Sciences, PHENIKAA University, Hanoi 12116, Vietnam}

\emailAdd{nhung.daothi@phenikaa-uni.edu.vn}
\emailAdd{ninh.leduc@phenikaa-uni.edu.vn}

\abstract{
We present new results for the theoretical prediction of doubly-polarized cross sections of 
$WZ$ events at the LHC using leptonic decays. Compared to the previous studies, two new kinematic cuts are considered. 
These cuts are designed to enhance the doubly-longitudinal (LL) polarization and, at the same time, 
study the Radiation Amplitude Zero effect. We found a new cut on the rapidity separation between the 
$Z$ boson and the electron from the $W$ decay which makes the LL fraction largest, namely $|\Delta y_{Z,e}| < 0.5$. 
This result is obtained at the next-to-leading order in the strong and electroweak couplings. 
}

\begin{document}
\maketitle
\flushbottom

\section{Introduction}
\label{sect:intro}
With the new results from ATLAS \cite{ATLAS:2022oge} where doubly-polarized cross sections 
of the diboson $W^\pm Z$ production at the Large Hadron Collider (LHC) are measured for the first time using 
the Run-2 data set and that the LHC Run 3 already began in July 2022, there is a foundation to expect that more precise measurements of diboson joint-polarization cross sections from ATLAS and CMS will come soon. 

Measuring the doubly-polarized cross sections in diboson production processes allows for testing the Standard Model (SM) 
at a deeper level as well as finding possible new physics effects via polarization observables. 
Recent theoretical works to define the signal part of the doubly-polarized cross sections using the double-pole 
approximation (DPA) were able to provide results at the next-to-leading-order (NLO) in QCD for $W^+W^-$ \cite{Denner:2020bcz}, 
$WZ$ \cite{Denner:2020eck,Le:2022lrp,Le:2022ppa,Denner:2022riz}, $ZZ$ \cite{Denner:2021csi} and at NLO in the electroweak (EW) 
interactions for $ZZ$ \cite{Denner:2021csi} and $WZ$ \cite{Le:2022lrp,Le:2022ppa}. For the case of $W^+W^-$, 
the next-to-next-to-leading-order QCD results are available \cite{Poncelet:2021jmj}.  

In our previous works \cite{Le:2022lrp,Le:2022ppa} the momenta of the final-state leptons are selected according to the ATLAS 
fiducial phase-space cut (named Cut 1 in this paper) as defined in \refs{Aaboud:2016yus,ATLAS:2019bsc,ATLAS:2022oge}. After discussions with experimental colleagues, we realized that other phase-space cuts (Cut 2 and Cut 3 in this paper) should be explored as well.  
These new kinematic setups are designed to enhance the doubly longitudinal (LL) polarization \cite{Franceschini:2017xkh} and 
observe the Radiation Amplitude Zero (RAZ) effect \cite{Baur:1994ia}.  
The purpose of this work is to explore these setups in the hope of finding optimal features for the study of the LL polarization 
in $WZ$ events. 

The paper is organized as follows. We first define the polarized cross sections in \sect{sect:pol_def}, before 
presenting the numerical results in \sect{sect:results}. Conclusions are provided in \sect{sect:conclusion}.  
%=========================================================================
\section{Definition of polarized cross sections}
\label{sect:pol_def}
\begin{figure}[ht!]
  \centering
  \includegraphics[height=5cm]{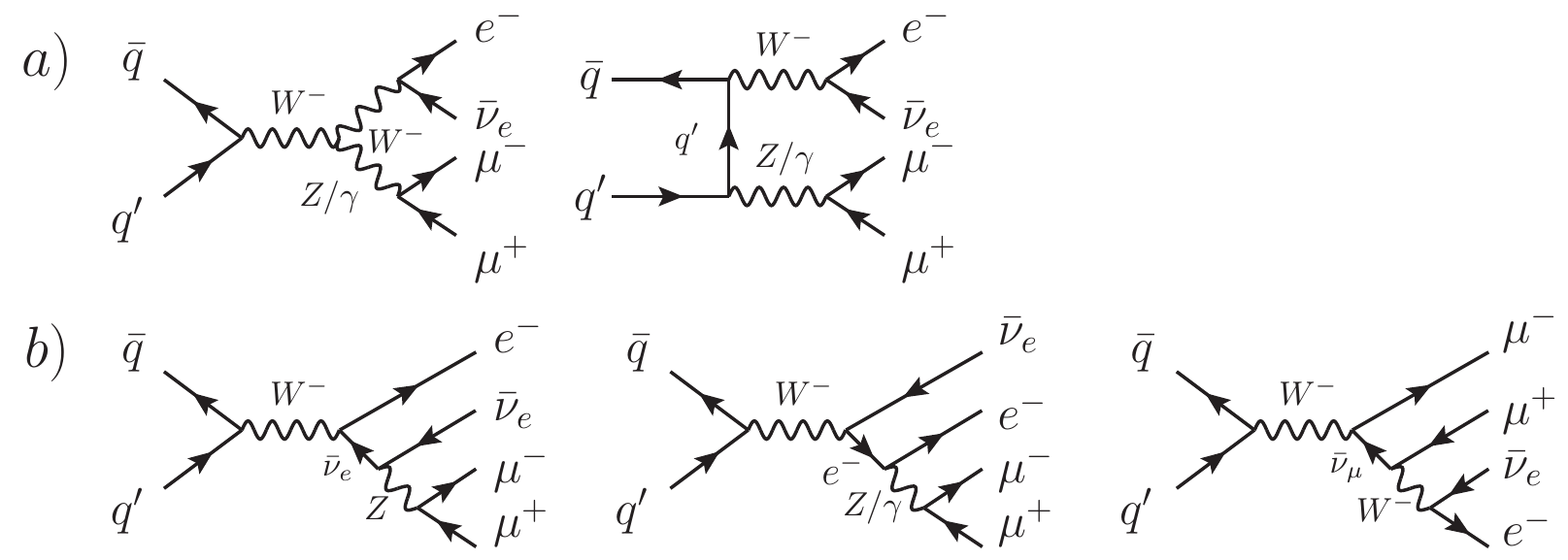}
  \caption{Double and single resonant diagrams at leading order. Group
    a) includes both double ($WZ$) and single ($W\gamma$) resonant diagrams, while group
    b) is only single resonant.}
  \label{fig:LO_diags}
\end{figure}

In order to set up our notations, we briefly 
review here the definition of the polarized cross sections. We use the same
conventions and calculation setup as in \cite{Le:2022ppa}.
The process of interest, which is measured at the LHC, reads
\bea
p(k_1) + p(k_2) \to \ell_1(k_3) + \ell_2(k_4) + \ell_3
(k_5) + \ell_4(k_6) + X,
\label{eq:proc1}
\eea
where the final-state leptons can be either $e^+\nu_e\mu^+\mu^-$ or 
$e^-\bar{\nu}_e\mu^+\mu^-$. Representative Feynman diagrams at leading
order (LO) are depicted in \fig{fig:LO_diags}.
%---begin: plots%
% ---end: plots%

From \fig{fig:LO_diags} we see that the doubly-polarized $WZ$ events, which occur via 
the double-resonant $WZ$ diagrams in the group a), are mixed with the single resonant events. Theoretically, 
we cannot just select those $WZ$ double resonant diagrams because they are linked with 
the single resonant diagrams by gauge invariance. To separate the $WZ$ events we need to use 
a special technique called the double-pole approximation, which selects only the gauge-invariance part 
of the $WZ$ double resonant cross section. We note that the DPA has been widely used in diboson production 
processes, see \bib{Denner:2000bj} and references therein.

To be more concrete, the $WZ$ double resonant processes are written as
\bea
p(k_1) + p(k_2) \to V_1(q_1) + V_2(q_2) \to \ell_1(k_3) + \ell_2(k_4) + \ell_3
(k_5) + \ell_4(k_6) + X,
\label{eq:proc1_VV}
\eea
where the intermediate gauge bosons are $V_1=W^\pm$, $V_2=Z$. 
Hence the double-pole unpolarized amplitude at leading
order (LO) can be expressed as
\bea
\mathcal{A}_\text{LO,DPA}^{\bar{q}q'\to V_1V_2\to 4l} = \fr{1}{Q_1Q_2}
\sum_{\lambda_1,\lambda_2=1}^{3}
\mathcal{A}_\text{LO}^{\bar{q}q'\to V_1V_2}(\hat{k}_i)\mathcal{A}_\text{LO}^{V_1\to
    \ell_1\ell_2}(\hat{k}_i)\mathcal{A}_\text{LO}^{V_2\to \ell_3\ell_4}(\hat{k}_i)
,\label{eq:LO_DPA}
\eea
with 
\bea
Q_j = q_j^2 - M_{V_j}^2 + iM_{V_j}\Gamma_{V_j}\, (j=1,2),
\label{eq:Qi_def}
\eea
where $q_1 = k_3+k_4$, $q_2 = k_5 + k_6$, $M_V$ and $\Gamma_V$ are the
physical mass and width of the gauge boson $V$, and $\lambda_j$ are the
polarization indices of the gauge bosons. 
Note that the helicity indices of the initial quarks and final leptons are implicit.
It is crucial that all helicity amplitudes $\mathcal{A}$ in
the r.h.s. are calculated using on-shell (OS) momenta $\hat{k}_i$ for the
final-state leptons as well as OS momenta $\hat{q}_j$ for the
intermediate gauge bosons, derived from the off-shell (full process)
momenta $k_i$ and $q_j$, in order to ensure that gauge invariance in
the amplitudes is preserved. An OS mapping is used to obtain the OS
momenta $\hat{k}_i$ from the off-shell momenta $k_i$. 
This OS mapping is not unique, however the shift induced by different mappings is of order
$\alpha \Gamma_V/(\pi M_V)$ \cite{Denner:2000bj}. The OS mapping used 
in this paper is the same as in \bib{Le:2022ppa}.

\eq{eq:LO_DPA} serves as the master equation to define the doubly polarized cross sections. 
Since a massive gauge boson has three
physical polarization states: two transverse states $\lambda = 1$ and
$\lambda = 3$ (left and right) and one longitudinal state
$\lambda = 2$, the $WZ$ system has in total $9$ polarization states. The unpolarized 
amplitude defined in \eq{eq:LO_DPA} is the sum of these $9$ polarized amplitudes. 
The unpolarized cross section is then divided into the following five terms:
\begin{itemize}
\item $W_L Z_L$: The longitudinal-longitudinal (LL) contribution, obtained with selecting $\lambda_1=\lambda_2=2$
  in the sum of \eq{eq:LO_DPA}; 
\item $W_L Z_T$: The longitudinal-transverse (LT) contribution, obtained with selecting $\lambda_1=2$, 
$\lambda_2=1,3$. The LT cross section includes the interference term between the $(21)$ and $(23)$ amplitudes.
\item $W_T Z_L$: The transverse-longitudinal (TL) contribution, obtained with selecting $\lambda_1=1,3$, 
$\lambda_2=2$. The interference between the $(12)$ and $(32)$ amplitudes is here included.
\item $W_T Z_T$: The transverse-transverse (TT) contribution, obtained with selecting $\lambda_1=1,3$, 
$\lambda_2=1,3$. The interference terms between the $(11)$, $(13)$, $(31)$, $(33)$ amplitudes are here included.
\item Interference: This includes the interference terms between the above LL, LT, TL, TT amplitudes.     
\end{itemize}
Our doubly-polarized cross-section results include not only the leading order but also the NLO QCD and EW corrections. The expressions for double-pole unpolarized amplitudes
 need to be extended to include also the virtual corrections, the gluon/photon induced and radiation processes as done in \bib{Le:2022ppa}. In this short writing the above LO definition of the polarized cross sections 
is enough for the reader to understand the numerical results discussed in the next section.
%=========================================================================
\section{Numerical results}
\label{sect:results}
Our numerical results are obtained for proton-proton collisions at a center-of-mass energy of $13$~TeV.  
Fixed factorization and renormalization scales are used, namely 
$\mu_F = \mu_R = \mu_0 = (M_W + M_Z)/2$, where $M_W=80.385\,\gev$ and $M_Z = 91.1876\,\gev$. 
For the parton distribution functions (PDF) and value of the strong coupling constant, 
the Hessian set 
{\tt
  LUXqed17\char`_plus\char`_PDF4LHC15\char`_nnlo\char`_30}~\bibs{Watt:2012tq,Gao:2013bia,Harland-Lang:2014zoa,Ball:2014uwa,Butterworth:2015oua,Dulat:2015mca,deFlorian:2015ujt,Carrazza:2015aoa,Manohar:2016nzj,Manohar:2017eqh} via the library {\tt LHAPDF6}~\bibs{Buckley:2014ana} is employed. More details about other input parameters are provided in \bib{Le:2022ppa}. 

For NLO EW corrections, an additional photon 
can be emitted. Hence, the lepton-photon recombination to define a dressed lepton is done before applying the
analysis cuts. A dressed lepton has the momentum 
of $p'_\ell = p_\ell + p_\gamma$ if the angular distance $\Delta
R(\ell,\gamma) \equiv \sqrt{(\Delta\eta)^2+(\Delta\phi)^2}< 0.1$, i.e. when the photon 
is close enough to the bare lepton. Here the letter $\ell$ denotes $e$ or $\mu$ 
and all momenta are calculated in the Lab frame. 
All leptons and quarks except for the top quark are approximated as massless.

The doubly polarized cross sections and distributions depend on the reference frame. 
We choose the $WZ$ center-of-mass frame, the same as in the ATLAS
measurement \cite{ATLAS:2022oge}. We now specify the three cut setups used in this paper. 
They read as follows.
\paragraph{Cut 1:} The baseline setup, called Cut 1, is the ATLAS fiducial set of cuts used in 
\refs{Aaboud:2016yus,ATLAS:2019bsc,ATLAS:2022oge}, which reads  
\bea
        p_{T,e} > 20\gev, \quad p_{T,\mu^\pm} > 15\gev, \quad |\eta_\ell|<2.5,\crn
        \Delta R\left(\mu^+,\mu^-\right) > 0.2, \quad \Delta R\left(e,\mu^\pm\right) > 0.3, \label{eq:cut_default}\\
        m_{T,W} > 30\gev, \quad \left|m_{\mu^+\mu^-} - M_Z\right| < 10\gev\, ,\nn
\eea 
where $m_{T,W} = \sqrt{2p_{T,\nu} p_{T,e} [1-\cos\Delta\phi(e,\nu)]}$ with $\Delta\phi(e,\nu)$ being 
the angle between the electron and the neutrino in the transverse plane. This Cut 1 was used in our previous studies \cite{Le:2022lrp,Le:2022ppa}.

\paragraph{Cut 2:} In addition to the cuts in Cut 1, we further require that the transverse momentum
of the $WZ$ system satisfies \cite{Franceschini:2017xkh}
\bea
p_{T,WZ} < 70\gev.
\label{eq:setup2}
\eea
At NLO, this additional cut affects only the real-emission contributions with an extra particle in the final state. 
The LO term and virtual corrections are unaffected.
The purpose of this cut is to observe the RAZ in the TT component, which is smeared out by QCD radiation \cite{Baur:1994ia,Franceschini:2017xkh}.  
This cut reduces higher-order QCD corrections. 
Dominant backgrounds, in particular $t\bar{t}$, $t\bar{t}V$, $VVV$ are expected to decrease significantly by this cut as well.  

\paragraph{Cut 3:} In addition to the cuts in Cut 2, we further require \cite{Franceschini:2017xkh}
\bea
p_{T,Z} > 200\gev.
\label{eq:setup3}
\eea
This additional cut reduces drastically the LO contribution as well as all NLO corrections. 
The purpose of this cut is to focus more on the high energy regime where new physics effects are expected to be present. 
As will be seen, this cut will increase the LL fraction significantly. 

\subsection{Integrated polarized cross sections}
\label{sect:XS}
%%%
\begin{table}[h!]
 \renewcommand{\arraystretch}{1.3}
\begin{bigcenter}
%   \small
%    \footnotesize
\setlength\tabcolsep{0.03cm}
\fontsize{7.0}{7.0}
\begin{tabular}{|c|c|c|c|c|c|c|c|c|c|}\hline
  & $\sigma_\text{LO}\,\text{[fb]}$ & $f_\text{LO}\,\text{[\%]}$  & $\sigma^\text{EW}_\text{NLO}\,\text{[fb]}$ & $f^\text{EW}_\text{NLO}\,\text{[\%]}$ & $\sigma^\text{QCD}_\text{NLO}\,\text{[fb]}$ & $f^\text{QCD}_\text{NLO}\,\text{[\%]}$ & $\sigma^\text{QCDEW}_\text{NLO}\,\text{[fb]}$ & $f^\text{QCDEW}_\text{NLO}\,\text{[\%]}$ & $\bar{\delta}_\text{EW}\,\text{[\%]}$\\
\hline
{\fontsize{6.0}{6.0}$\text{Unpol., Cut 1}$} & $18.934(1)^{+4.8\%}_{-5.9\%}$ & $100$ & $18.138(1)^{+4.9\%}_{-6.0\%}$ & $100$ & $34.071(2)^{+5.3\%}_{-4.2\%}$ & $100$ & $33.275(2)^{+5.4\%}_{-4.3\%}$ & $100$ & $-2.3$\\
{\fontsize{6.0}{6.0}$\text{Cut 2}$} & $18.934(1)^{+4.8\%}_{-5.9\%}$ & $100$ & $17.897(1)^{+4.9\%}_{-6.0\%}$ & $100$ & $25.860(3)^{+3.2\%}_{-2.5\%}$ & $100$ & $24.823(3)^{+3.4\%}_{-2.6\%}$ & $100$ & $-4.0$\\
{\fontsize{6.0}{6.0}$\text{Cut 3}$} & $0.392^{+1.4\%}_{-1.8\%}$     & $100$ & $0.343_{-1.5\%}^{+1.1\%}$     & $100$ & $0.445_{-1.5\%}^{+2.2\%}$     & $100$ & $0.396_{-1.2\%}^{+2.0\%}$     & $100$ & $-11.0$\\ 
\hline
{\fontsize{6.0}{6.0}$W^+_{L}Z_{L}$, $\text{Cut 1}$} & $1.492^{+5.1\%}_{-6.3\%}$ & $7.9$ & $1.428^{+5.2\%}_{-6.4\%}$ & $7.9$ & $1.938^{+2.7\%}_{-2.2\%}$ & $5.7$ & $1.874^{+2.8\%}_{-2.3\%}$ & $5.6$ & $-3.3$\\
{\fontsize{6.0}{6.0}$\text{Cut 2}$} & $1.492^{+5.1\%}_{-6.3\%}$ & $7.9$ & $1.420^{+5.3\%}_{-6.4\%}$ & $7.9$ & $1.786^{+1.9\%}_{-2.3\%}$ & $6.9$ & $1.714^{+2.0\%}_{-2.2\%}$ & $6.9$ & $-4.0$\\
{\fontsize{6.0}{6.0}$\text{Cut 3}$} & $0.105_{-0.7\%}^{+0.0\%}$ & $26.7$ & $0.092_{-0.5\%}^{+0.0\%}$ & $26.9$ & $0.100_{-0.6\%}^{+0.8\%}$ & $22.6$ & $0.088_{-1.0\%}^{+1.2\%}$ & $22.2$ & $-13.0$\\
\hline
{\fontsize{6.0}{6.0}$W^+_{L}Z_{T}$, $\text{Cut 1}$} & $2.018^{+5.8\%}_{-7.0\%}$ & $10.7$ & $1.951^{+5.8\%}_{-7.0\%}$ & $10.8$ & $5.273^{+7.3\%}_{-5.9\%}$ & $15.5$ & $5.207^{+7.4\%}_{-6.0\%}$ & $15.6$ & $-1.3$\\
{\fontsize{6.0}{6.0}$\text{Cut 2}$} & $2.018^{+5.8\%}_{-7.0\%}$ & $10.7$ & $1.928^{+5.8\%}_{-7.0\%}$ & $10.8$ & $3.419^{+4.9\%}_{-3.8\%}$ & $13.2$ & $3.329^{+5.1\%}_{-3.9\%}$ & $13.4$ & $-2.6$\\
{\fontsize{6.0}{6.0}$\text{Cut 3}$} & $0.017_{-0.4\%}^{+0.0\%}$ & $4.4$ & $0.016_{-0.5\%}^{+0.0\%}$ & $4.8$ & $0.023_{-2.9\%}^{+3.9\%}$ & $5.1$ & $0.022_{-2.9\%}^{+3.9\%}$ & $5.5$ & $-4.3$\\
\hline
{\fontsize{6.0}{6.0}$W^+_{T}Z_{L}$, $\text{Cut 1}$} & $1.903^{+5.7\%}_{-6.9\%}$ & $10.1$ & $1.893^{+5.7\%}_{-6.9\%}$ & $10.4$ & $5.024^{+7.4\%}_{-5.9\%}$ & $14.7$ & $5.013^{+7.4\%}_{-5.9\%}$ & $15.1$ & $-0.2$\\
{\fontsize{6.0}{6.0}$\text{Cut 2}$} & $1.903^{+5.7\%}_{-6.9\%}$ & $10.1$ & $1.826^{+5.8\%}_{-7.0\%}$ & $10.2$ & $3.281^{+5.0\%}_{-4.0\%}$ & $12.7$ & $3.204^{+5.1\%}_{-4.1\%}$ & $12.9$ & $-2.3$\\
{\fontsize{6.0}{6.0}$\text{Cut 3}$} & $0.017_{-0.4\%}^{+0.0\%}$ & $4.3$ & $0.017_{-0.5\%}^{+0.0\%}$ & $4.9$ & $0.021_{-2.5\%}^{+3.1\%}$ & $4.6$ & $0.020_{-2.3\%}^{+3.0\%}$ & $5.2$ & $0.0$\\
\hline
{\fontsize{6.0}{6.0}$W^+_{T}Z_{T}$, $\text{Cut 1}$} & $13.376^{+4.5\%}_{-5.6\%}$ & $70.6$ & $12.728(1)^{+4.6\%}_{-5.7\%}$ & $70.2$ & $21.626(2)^{+4.5\%}_{-3.6\%}$ & $63.5$ & $20.977(2)^{+4.7\%}_{-3.8\%}$ & $63.0$ & $-3.0$\\
{\fontsize{6.0}{6.0}$\text{Cut 2}$} & $13.376^{+4.5\%}_{-5.6\%}$ & $70.6$ & $12.587(1)^{+4.5\%}_{-5.7\%}$ & $70.3$ & $17.132(2)^{+2.6\%}_{-2.1\%}$ & $66.2$ & $16.342(2)^{+2.7\%}_{-2.2\%}$ & $65.8$ & $-4.6$\\
{\fontsize{6.0}{6.0}$\text{Cut 3}$} & $0.247_{-2.5\%}^{+2.2\%}$ & $63.1$ & $0.212_{-2.1\%}^{+1.9\%}$ & $61.9$ & $0.297_{-2.1\%}^{+3.1\%}$ & $66.7$ & $0.262_{-1.7\%}^{+3.0\%}$ & $66.1$ & $-11.8$\\
\hline
{\fontsize{6.0}{6.0}$\text{Inter.}$, $\text{Cut 1}$} & $0.144(1)$ & $0.8$ & $0.138(1)$ & $0.8$ & $0.210(3)$ & $0.6$ & $0.204(3)$ & $0.6$ & $-2.9$\\
{\fontsize{6.0}{6.0}$\text{Cut 2}$} & $0.144(1)$ & $0.8$ & $0.137(1)$ & $0.8$ & $0.242(3)$ & $0.9$ & $0.235(3)$ & $0.9$ & $-2.9$\\
{\fontsize{6.0}{6.0}$\text{Cut 3}$} & $0.006$    & $1.5$ & $0.005$    & $1.6$ & $0.005$    & $1.0$ & $0.004$    & $1.0$ & $-20.0$\\
\hline
\end{tabular}
%%%
\caption{\small Unpolarized (Unpol.) and doubly polarized cross sections in fb
  together with polarization fractions calculated at LO, NLO EW, NLO
  QCD, and NLO QCD+EW, all in the DPA, in the $WZ$ center-of-mass system for the
  $W^+ Z$ process. The interference (Inter.) between the polarized amplitudes 
  is provided in the bottom row.  
  The statistical uncertainties (in parenthesis) are given on the last
  digits of the central prediction when significant. Seven-point scale
  uncertainty is also provided for the cross sections as sub- and
  superscripts in percent. In the last column the EW correction relative to the NLO QCD prediction 
  is given.}
\label{tab:xs_fr_Wp}
\end{bigcenter}
\end{table}
%%%
\begin{table}[h!]
 \renewcommand{\arraystretch}{1.3}
\begin{bigcenter}
%   \small
%    \footnotesize
\setlength\tabcolsep{0.03cm}
\fontsize{7.0}{7.0}
\begin{tabular}{|c|c|c|c|c|c|c|c|c|c|}\hline
  & $\sigma_\text{LO}\,\text{[fb]}$ & $f_\text{LO}\,\text{[\%]}$  & $\sigma^\text{EW}_\text{NLO}\,\text{[fb]}$ & $f^\text{EW}_\text{NLO}\,\text{[\%]}$ & $\sigma^\text{QCD}_\text{NLO}\,\text{[fb]}$ & $f^\text{QCD}_\text{NLO}\,\text{[\%]}$ & $\sigma^\text{QCDEW}_\text{NLO}\,\text{[fb]}$ & $f^\text{QCDEW}_\text{NLO}\,\text{[\%]}$ & $\bar{\delta}_\text{EW}\,\text{[\%]}$\\
\hline
{\fontsize{6.0}{6.0}$\text{Unpol., Cut 1}$} & $12.745^{+4.9\%}_{-6.2\%}$ & $100$ 
& $12.224^{+5.1\%}_{-6.3\%}$ & $100$ 
& $23.705(1)^{+5.5\%}_{-4.4\%}$ & $100$ 
& $23.184(1)^{+5.6\%}_{-4.5\%}$ & $100$ & $-2.2$\\
{\fontsize{6.0}{6.0}$\text{Cut 2}$} & $12.745^{+4.9\%}_{-6.2\%}$ & $100$ 
& $12.060(1)_{-6.3\%}^{+5.1\%}$ & $100$ 
& $17.905(2)_{-2.7\%}^{+3.4\%}$ & $100$ 
& $17.221(2)_{-2.8\%}^{+3.5\%}$ & $100$ & $-3.8$\\
{\fontsize{6.0}{6.0}$\text{Cut 3}$} & $0.209_{-1.8\%}^{+1.3\%}$ & $100$ 
& $0.184_{-1.5\%}^{+1.0\%}$ & $100$ 
& $0.259_{-2.7\%}^{+3.3\%}$ & $100$ 
& $0.234_{-2.5\%}^{+3.2\%}$ & $100$ & $-9.7$\\
\hline
{\fontsize{6.0}{6.0}$W^-_{L}Z_{L}$, $\text{Cut 1}$} & $1.094^{+5.2\%}_{-6.5\%}$ & $8.6$ 
& $1.048^{+5.3\%}_{-6.6\%}$ & $8.6$ 
& $1.407^{+2.6\%}_{-2.1\%}$ & $5.9$ 
& $1.361^{+2.7\%}_{-2.2\%}$ & $5.9$ & $-3.3$\\
{\fontsize{6.0}{6.0}$\text{Cut 2}$} & $1.094^{+5.2\%}_{-6.5\%}$ & $8.6$ 
& $1.043_{-6.6\%}^{+5.3\%}$ & $8.6$ 
& $1.308_{-2.3\%}^{+1.9\%}$ & $7.3$ 
& $1.257_{-2.3\%}^{+2.0\%}$ & $7.3$ & $-3.9$\\
{\fontsize{6.0}{6.0}$\text{Cut 3}$} & $0.059_{-0.8\%}^{+0.1\%}$ & $28.4$ 
& $0.052_{-0.6\%}^{+0.0\%}$ & $28.5$ 
& $0.057_{-0.4\%}^{+0.7\%}$ & $22.2$ 
& $0.051_{-0.7\%}^{+1.1\%}$ & $21.6$ & $-12.3$\\
\hline
{\fontsize{6.0}{6.0}$W^-_{L}Z_{T}$, $\text{Cut 1}$} & $1.508^{+5.8\%}_{-7.0\%}$ & $11.8$ 
& $1.456^{+5.9\%}_{-7.1\%}$ & $11.9$ 
& $3.921^{+7.3\%}_{-5.9\%}$ & $16.5$ 
& $3.869^{+7.4\%}_{-6.0\%}$ & $16.7$ & $-1.3$\\
{\fontsize{6.0}{6.0}$\text{Cut 2}$} & $1.508^{+5.8\%}_{-7.0\%}$ & $11.8$ 
& $1.440_{-7.1\%}^{+5.8\%}$ & $11.9$ 
& $2.605_{-4.0\%}^{+5.0\%}$ & $14.5$ 
& $2.536_{-4.1\%}^{+5.2\%}$ & $14.7$ & $-2.6$\\
{\fontsize{6.0}{6.0}$\text{Cut 3}$} & $0.010_{-0.5\%}^{+0.0}$ & $4.8$ 
& $0.010_{-0.4\%}^{+0.0\%}$ & $5.2$ 
& $0.015_{-4.2\%}^{+5.3\%}$ & $5.8$ 
& $0.014_{-4.2\%}^{+5.4\%}$ & $6.2$ & $0.0$\\
\hline
{\fontsize{6.0}{6.0}$W^-_{T}Z_{L}$, $\text{Cut 1}$} & $1.356^{+5.8\%}_{-7.0\%}$ & $10.6$ 
& $1.347^{+5.8\%}_{-7.0\%}$ & $11.0$ 
& $3.606^{+7.4\%}_{-6.0\%}$ & $15.2$ 
& $3.597^{+7.4\%}_{-6.0\%}$ & $15.5$ & $-0.2$\\
{\fontsize{6.0}{6.0}$\text{Cut 2}$} & $1.356^{+5.8\%}_{-7.0\%}$ & $10.6$ 
& $1.302_{-7.1\%}^{+5.9\%}$ & $10.8$ 
& $2.375_{-4.1\%}^{+5.1\%}$ & $13.3$ 
& $2.322_{-4.2\%}^{+5.2\%}$ & $13.5$ & $-2.3$\\
{\fontsize{6.0}{6.0}$\text{Cut 3}$} & $0.010_{-0.5\%}^{+0.0}$ & $4.7$ 
& $0.010_{-0.4\%}^{+0.0\%}$ & $5.2$ 
& $0.012_{-2.5\%}^{+3.9\%}$ & $4.7$ 
& $0.012_{-2.3\%}^{+3.7\%}$ & $5.1$ & $0.0$\\
\hline
{\fontsize{6.0}{6.0}$W^-_{T}Z_{T}$, $\text{Cut 1}$} & $8.833^{+4.6\%}_{-5.8\%}$ & $69.3$ 
& $8.416^{+4.8\%}_{-5.9\%}$ & $68.8$ 
& $14.664(1)^{+4.7\%}_{-3.8\%}$ & $61.9$ 
& $14.247(1)^{+4.9\%}_{-3.9\%}$ & $61.5$ & $-2.8$\\
{\fontsize{6.0}{6.0}$\text{Cut 2}$} & $8.833^{+4.6\%}_{-5.8\%}$ & $69.3$ 
& $8.321_{-6.0\%}^{+4.8\%}$ & $69.0$ 
& $11.549(1)_{-2.2\%}^{+2.8\%}$ & $64.5$ 
& $11.037(1)_{-2.3\%}^{+2.9\%}$ & $64.1$ & $-4.4$\\
{\fontsize{6.0}{6.0}$\text{Cut 3}$} & $0.129_{-2.5\%}^{2.2\%}$ & $61.8$ 
& $0.111_{-2.2\%}^{+1.7\%}$ & $60.7$ 
& $0.174_{-3.6\%}^{+4.8\%}$ & $67.1$ 
& $0.156_{-3.5\%}^{+4.8\%}$ & $66.8$ & $-10.3$\\
\hline
{\fontsize{6.0}{6.0}$\text{Inter., Cut 1}$} & $-0.046(1)$ & $-0.4$ 
& $-0.043(1)$ & $-0.4$ 
& $0.107(2)$ & $0.5$ 
& $0.110(2)$ & $0.5$ & $+2.8$\\
{\fontsize{6.0}{6.0}$\text{Cut 2}$} & $-0.046(1)$ & $-0.4$ 
& $-0.045(1)$ & $-0.4$ & $0.068(2)$ & $0.4$ & $0.069(2)$ & $0.4$ & $+1.5$\\
{\fontsize{6.0}{6.0}$\text{Cut 3}$} & $0.001$ & $0.4$ & $0.001$ & $0.4$ & $0.001$ & $0.3$ & $0.001$ & $0.3$ & $0.0$\\
\hline
\end{tabular}
%%%
\caption{\small Same as \tab{tab:xs_fr_Wp} but for the 
 $W^- Z$ process.}
\label{tab:xs_fr_Wm}
\end{bigcenter}
\end{table}
%%%
We first present results for the integrated cross sections at LO, NLO QCD, NLO EW, and
NLO QCD+EW for the unpolarized case, LL, LT, TL, TT polarizations and the interference in \tab{tab:xs_fr_Wp} for the case of $W^+Z$ 
and in \tab{tab:xs_fr_Wm} for $W^-Z$ for all three cut setups. The results for Cut 1 have already been published 
in \refs{Le:2022lrp,Le:2022ppa}. They are re-provided here for the sake of comparison. 

Included in the two tables are also the polarization fractions, $f$, calculated as
ratios of the polarized cross sections over the unpolarized cross section at each level of
accuracy. The total EW correction relative to the NLO QCD prediction is defined as
\begin{equation} 
  \bar{\delta}_\text{EW} = (\sigma_\text{NLO}^\text{QCDEW} -
  \sigma_\text{NLO}^\text{QCD})/\sigma_\text{NLO}^\text{QCD}.
  \label{eq:deltaEW}
\end{equation} 
This information is shown in the last column.
Statistical errors are very small and shown in a few places where they are significant. 
Scale uncertainties are much bigger and are provided for the cross sections as sub- and 
superscripts in percent. These uncertainties are calculated using the seven-point method 
where the two scales $\mu_F$ and $\mu_R$ are varied as 
$n\mu_0/2$ with $n=1,2,4$ and $\mu_0
= (M_W + M_Z)/2$ being the central scale. Additional
constraint $1/2 \leq \mu_R/\mu_F \leq 2$ is used to limit the number of scale
choices to seven at NLO QCD. The cases $\mu_R/\mu_F = 1/4$ or $4$ are excluded, being considered too extreme. 

From the tables, we see that the veto cut of $p_{T,WZ} < 70$ GeV reduces the NLO QCD+EW unpolarized cross section by around $25.5\%$ for both processes, 
which is almost entirely due to the reduction of the QCD correction. This reduction is however not equally distributed among 
different polarizations. For the $W^+Z$ channel, they are $-8.5\%$, $-36.1$, $-36.1\%$, $-22.1\%$ for the $W_L Z_L$, $W_L Z_T$, 
$W_T Z_L$, $W_T Z_T$, respectively. The corresponding numbers for the $W^- Z$ case are $-7.6\%$, $-34.5\%$, 
$-35.4\%$, $-22.5\%$.  
One notices that, for both processes, the $LL$ component is least reduced while the mixed polarizations $LT$ 
and $TL$ are most affected by the veto cut. This is reflected in the polarization fractions.
The $f_{LL}$ is increased from $5.6\%$ ($5.9\%$) to $6.9\%$ ($7.3\%$) for the $W^+ Z$ ($W^- Z$) at NLO QCD+EW. 
Both $f_{LT}, f_{TL}$ decrease two percent, but the doubly-tranverse polarization fraction 
increases from $63.0\%$ ($61.5\%$) to $65.8\%$ ($64.1\%$).

Moving to Cut 3, we see that the integrated cross sections are drastically reduced, by around $99\%$ compared to Cut 1, for the unpolarized case and for both processes. With $139 \text{fb}^{-1}$ data, the numbers of signal events for Cut 1 are $1190$, $1900$, $3100$, $10900$ for the LL, LT, TL, TT polarizations, summing over the two processes, as shown in Table 1 (left) of \cite{ATLAS:2022oge}. 
Observing that the reduction is not uniform for different polarizations, 
the corresponding results for Cut 3 are obtained as $51$, $8$, $11$, $129$. These numbers will increased greatly when Run-3 data is added to the analysis in the near future.
The purpose of Cut 3 is to enhance the LL fraction and this can be seen clearly in the tables. The $f_{LL}$ now reads $22.2\%$ ($21.6\%$) at NLO QCD+EW for $W^+ Z$ ($W^- Z$) process, being ranked second after the TT fraction.     

Concerning the EW corrections, they are all negative (except for the interference) and 
their absolute values are all smaller than $5\%$ for Cut 1 and Cut 2. For Cut 3, they are greater 
than $10\%$ for the LL and TT cases, signifying the importance of the EW corrections for the future measurements of the LL fraction.
%%%
\subsection{Kinematic distributions}
\label{sect:XS}
\begin{figure}[h!]
  \centering
  \begin{tabular}{cc}
  \includegraphics[width=0.48\textwidth]{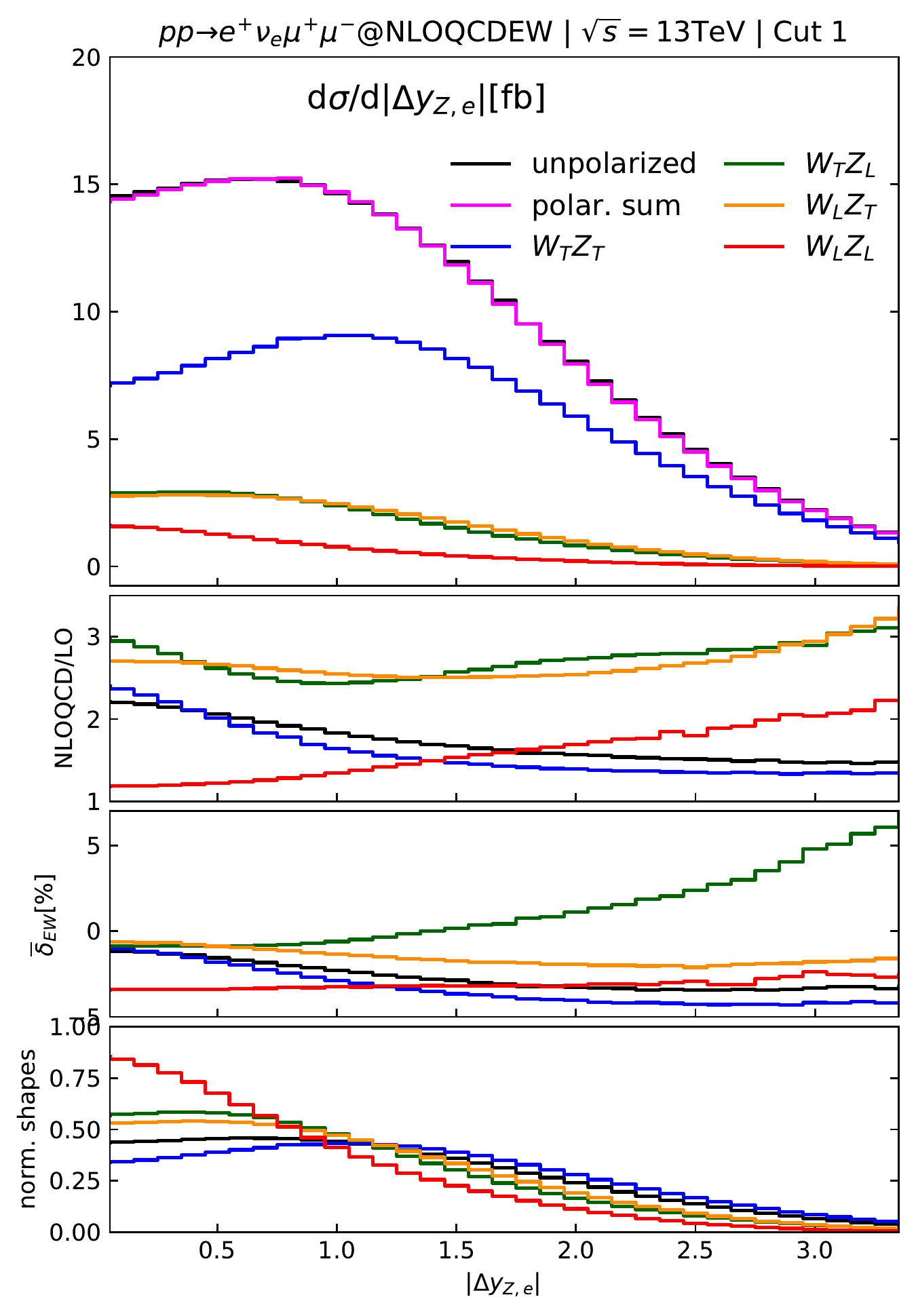} 
  \includegraphics[width=0.48\textwidth]{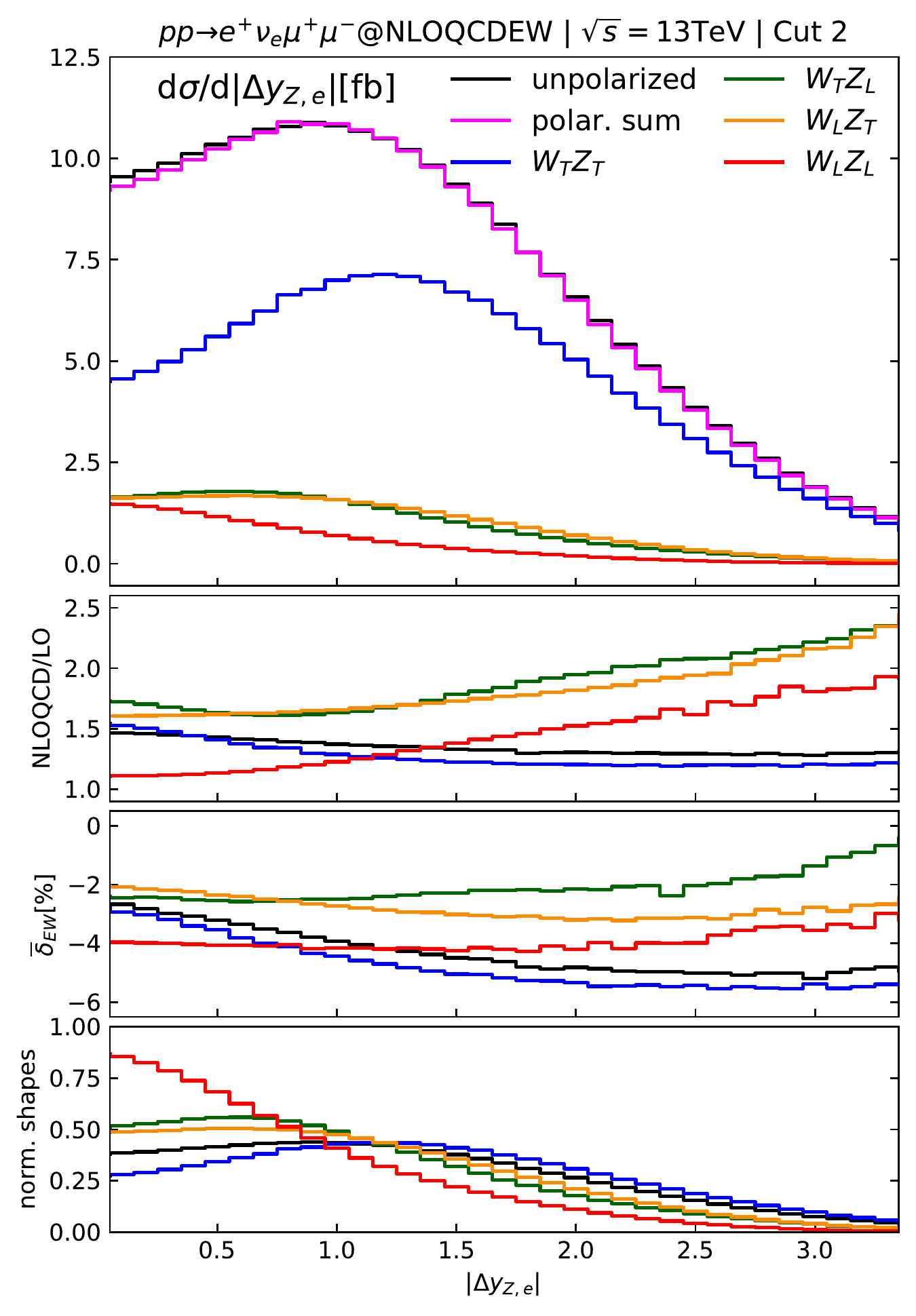}\\ 
  \includegraphics[width=0.48\textwidth]{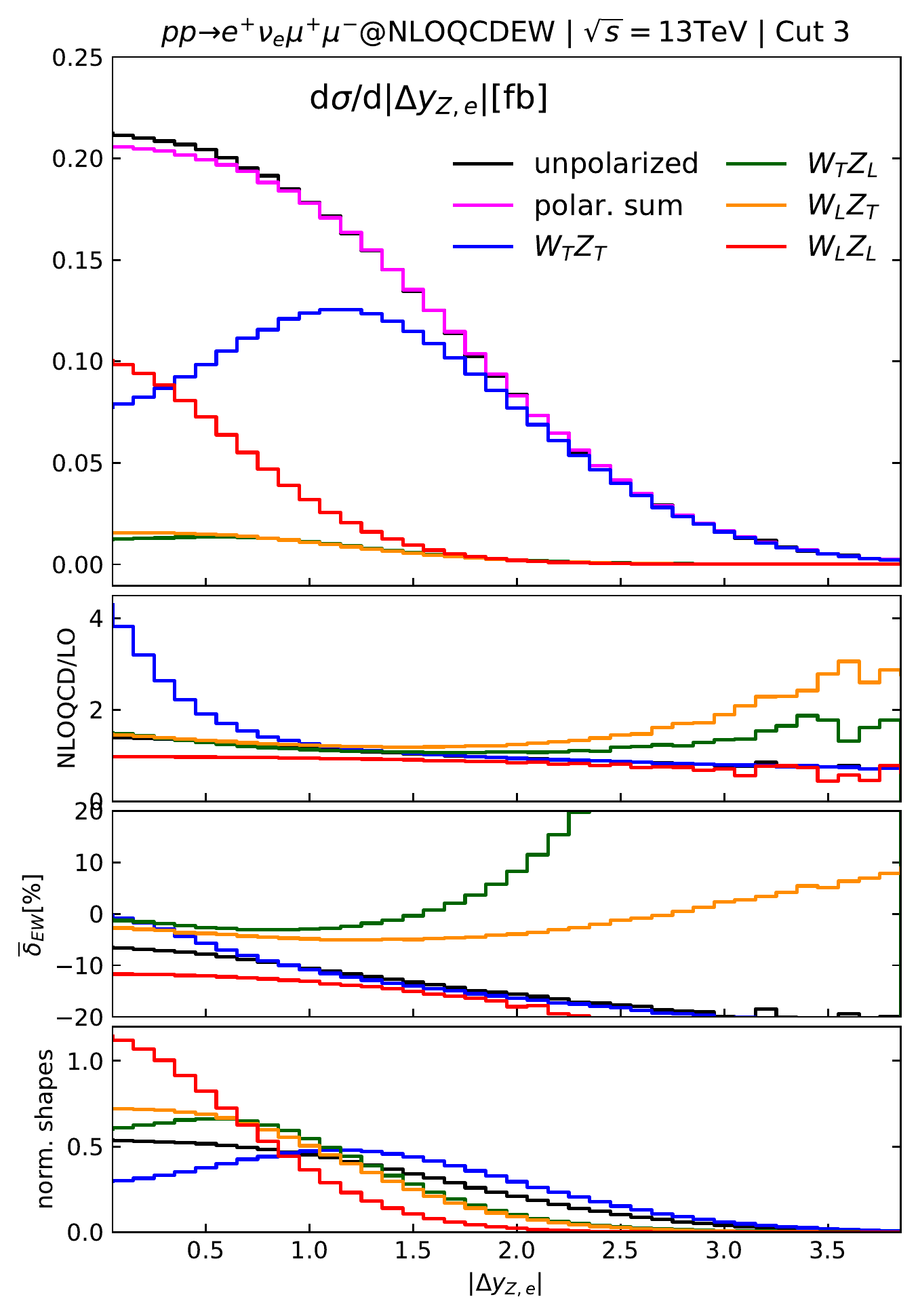}
  \end{tabular}
  \caption{\footnotesize Distributions in the rapidity
    separation (in absolute value) between the positron and
    the $Z$ boson for Cut 1 (top left), Cut 2 (top right), and Cut 3 (bottom) of the $W^+ Z$ process. 
    The big panel shows the absolute values of the cross sections at NLO QCD+EW. The
    middle-up panel displays the ratio of the NLO QCD cross sections
    to the corresponding LO ones. The middle-down panel shows
    $\bar{\delta}_{\text{EW}}$, the EW corrections relative to the NLO
    QCD cross sections, in percent. In the bottom panel, the
    normalized shapes of the distributions are plotted to highlight
    differences in shape.}
  \label{fig:dist_y_Ze_Wp}
\end{figure}

\begin{figure}[h!]
  \centering
  \begin{tabular}{cc}
  \includegraphics[width=0.48\textwidth]{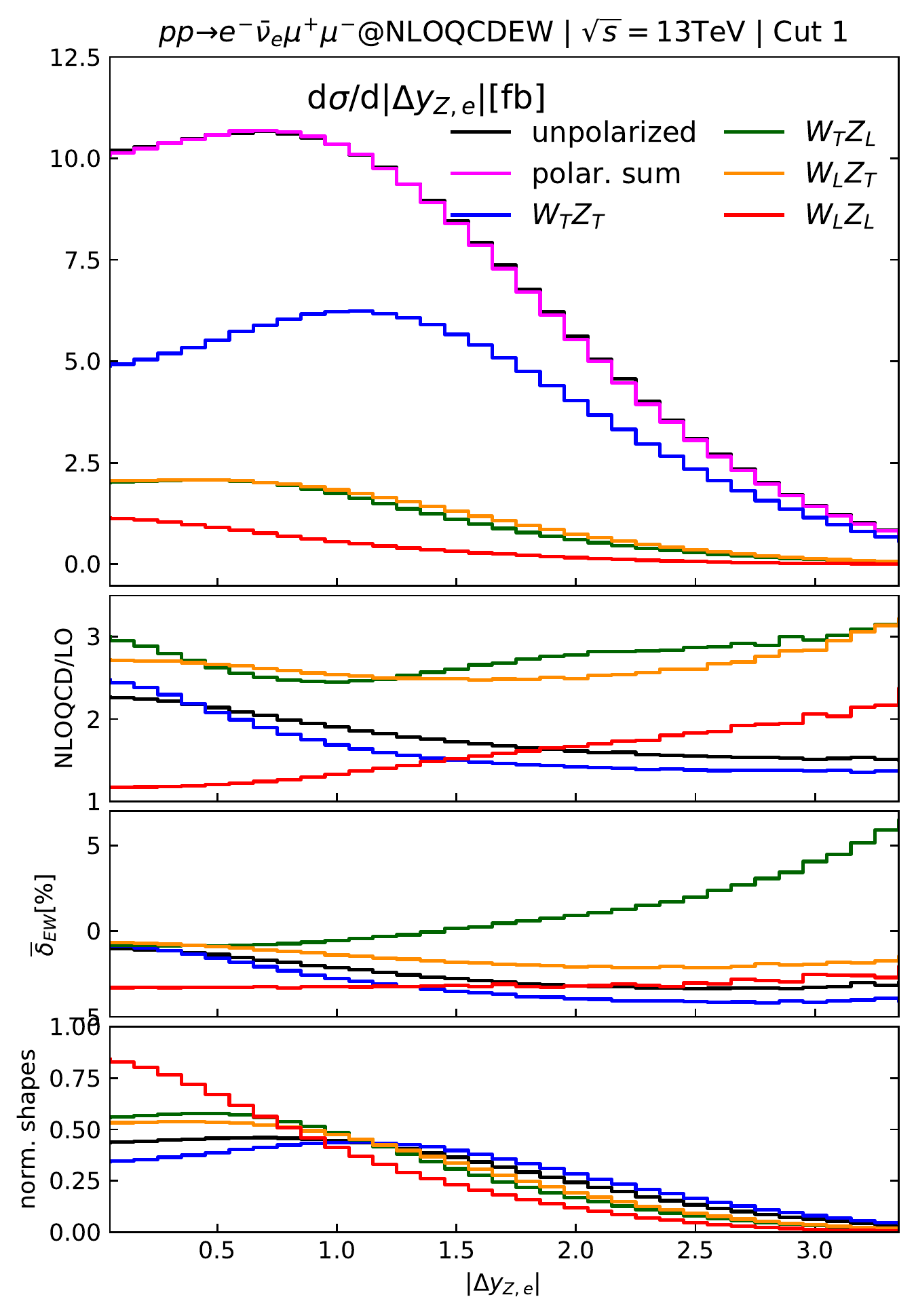}   
  \includegraphics[width=0.48\textwidth]{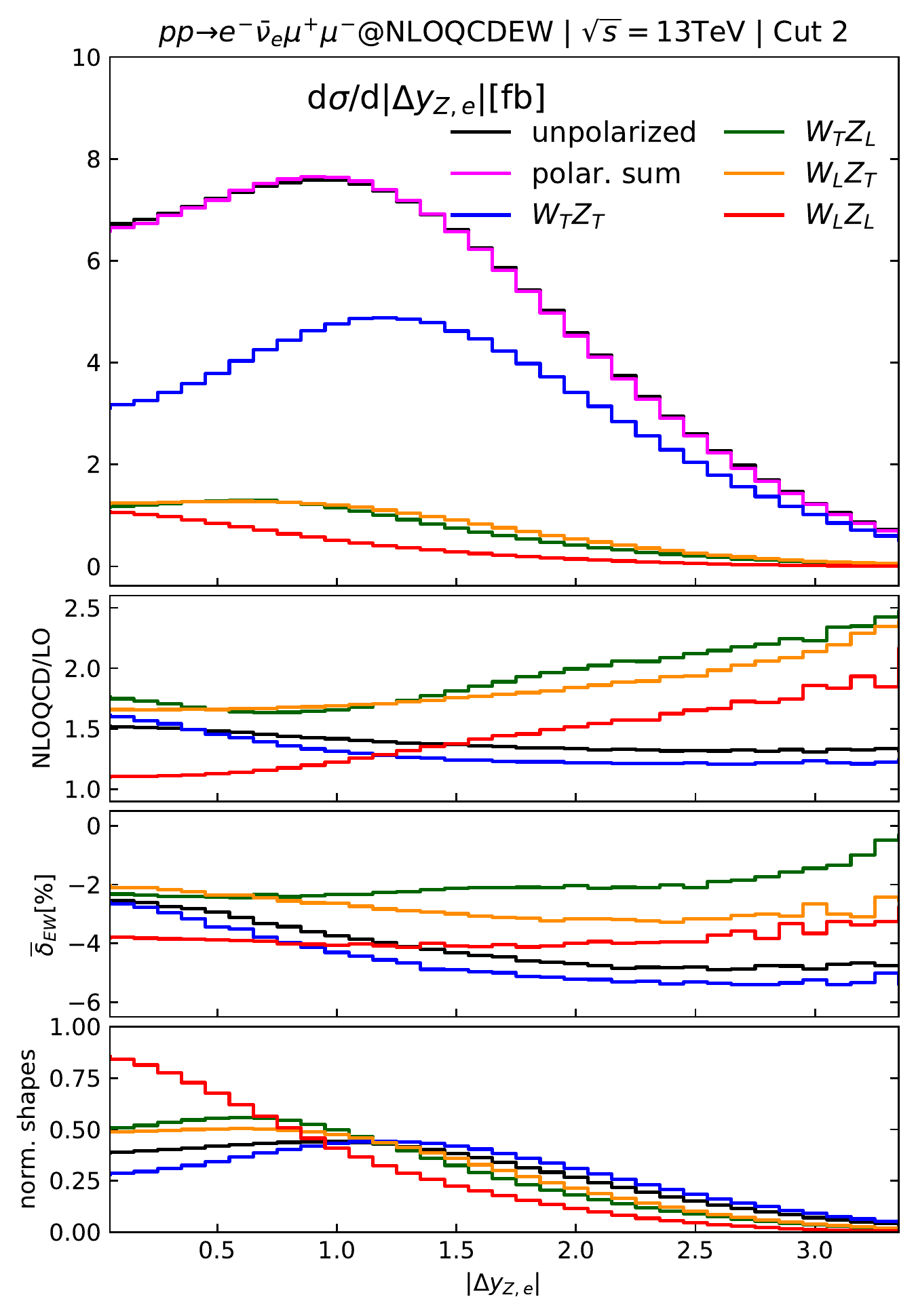}\\ 
  \includegraphics[width=0.48\textwidth]{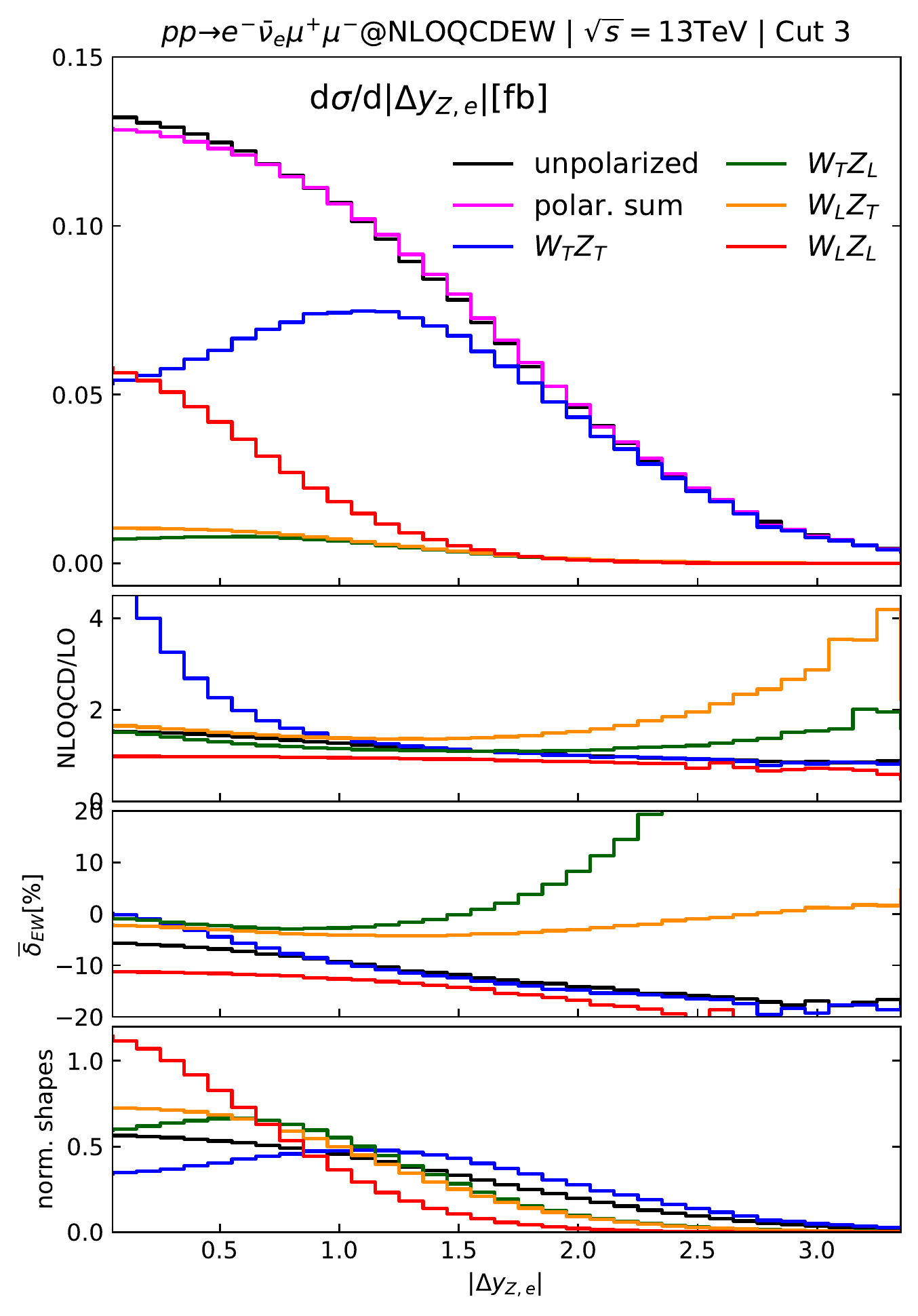}
  \end{tabular}
  \caption{Same as \fig{fig:dist_y_Ze_Wp} but for the 
 $W^- Z$ process.}
  \label{fig:dist_y_Ze_Wm}
\end{figure}

In \fig{fig:dist_y_Ze_Wp} ($W^+ Z$) and \fig{fig:dist_y_Ze_Wm} ($W^- Z$) we present new results for Cut 2 and Cut 3 on 
the distributions of rapidity separation between the positron (or electron) and the $Z$ boson directions. 
The plots for Cut 1, already shown in Fig.~5 of \bib{Le:2022ppa}, are displayed here for the sake of comparison. 

Comparing Cut 1 and Cut 2, we see that the above reduction in the cross section comes from the phase space region 
of $|\Delta y_{Z,e}| \approx 0$, where the TT, TL, and LT are most affected, while the LL changes slightly. 
The TL and LT cross sections are still larger than the LL one, but the difference is small. 

Moving to Cut 3, we find, very surprisingly, that the LL cross section is largest when the rapidity separation is smaller 
than $0.3$ for the $W^+ Z$ case. The value is $0.1$ for the $W^- Z$ process. 
Since the LL cross section is maximal at zero separation while the TT one is maximal at around $1.2$, we can further 
suppress the TT events by imposing an additional cut on the rapidity separation, namely $|\Delta y_{Z,e}| < \Delta y_\text{cut}$ 
with $\Delta y_\text{cut}$ in the range $[0.5,1]$. Results for various values of $\Delta y_\text{cut}$ are shown in 
\tab{tab:xs_Dy_Ze_Wpm}. There, in the parentheses, we also provide the acceptance $A$, defined as the ratio of the 
cross section after applying the $\Delta y_{Z,e}$ cut with respect to the one before applying this cut. 
From the table, we observe that the LL cross section is largest when $\Delta y_\text{cut}$ smaller than $0.6$ for the $W^+ Z$ process, 
and $0.4$ for the $W^- Z$ case. For the combined set of events, choosing $\Delta y_\text{cut} = 0.5$ guarantees that the 
LL fraction is dominant with an acceptance of $53\%$. We are confident that this can be done for the Run-3 data set.        
%%%
\clearpage
%%%
\clearpage
%%%
\clearpage
\begin{table}[th!]
 \renewcommand{\arraystretch}{1.3}
\begin{bigcenter}
%   \small
%    \footnotesize
\setlength\tabcolsep{0.03cm}
\fontsize{7.0}{7.0}
\begin{tabular}{|c|c|c|c|c|c|c|c|c|}\hline
 & \multicolumn{4}{c|}{$W^+ Z$} & \multicolumn{4}{c|}{$W^- Z$}\\
 \hline
 $\Delta y_\text{cut}$ & $\sigma_\text{TT}\,\text{[fb](A[\%])}$ & $\sigma_\text{LL}\,\text{[fb](A[\%])}$ & $\sigma_\text{LT}\,\text{[fb](A[\%])}$ & $\sigma_\text{TL}\,\text{[fb](A[\%])}$ & $\sigma_\text{TT}\,\text{[fb](A[\%])}$ & $\sigma_\text{LL}\,\text{[fb](A[\%])}$ & $\sigma_\text{LT}\,\text{[fb](A[\%])}$ & $\sigma_\text{TL}\,\text{[fb](A[\%])}$\\
\hline
$0.1$ & $0.008(3.0)$ & $0.010(11.4)$ & $0.002(7.2)$ & $0.001(6.0)$ & $0.005(3.4)$ & $0.006(11.4)$ & $0.001(7.2)$ & $0.001(6.0)$\\
$0.2$ & $0.016(6.0)$ & $0.020(22.6)$ & $0.003(14.4)$ & $0.002(12.1)$ & $0.011(6.9)$ & $0.011(22.6)$ & $0.002(14.5)$ & $0.001(12.0)$\\
$0.3$ & $0.024(9.1)$ & $0.029(33.3)$ & $0.005(21.6)$ & $0.004(18.4)$ & $0.016(10.4)$ & $0.017(33.3)$ & $0.003(21.7)$ & $0.002(18.2)$\\
$0.4$ & $0.033(12.4)$ & $0.038(43.3)$ & $0.006(28.7)$ & $0.005(24.7)$ & $0.022(14.1)$ & $0.022(43.3)$ & $0.004(28.8)$ & $0.003(24.6)$\\
$0.5$ & $0.042(16.0)$ & $0.046(52.5)$ & $0.008(35.7)$ & $0.006(31.3)$ & $0.028(18.0)$ & $0.027(52.5)$ & $0.005(35.8)$ & $0.004(31.1)$\\
$0.6$ & $0.052(19.7)$ & $0.053(60.7)$ & $0.009(42.6)$ & $0.008(37.9)$ & $0.034(22.1)$ & $0.031(60.8)$ & $0.006(42.7)$ & $0.005(37.7)$\\
$0.7$ & $0.062(23.7)$ & $0.060(68.0)$ & $0.011(49.3)$ & $0.009(44.5)$ & $0.041(26.3)$ & $0.034(68.0)$ & $0.007(49.3)$ & $0.005(44.3)$\\
$0.8$ & $0.073(28.0)$ & $0.065(74.2)$ & $0.012(55.6)$ & $0.010(51.0)$ & $0.048(30.8)$ & $0.038(74.3)$ & $0.008(55.6)$ & $0.006(50.9)$\\
$0.9$ & $0.085(32.4)$ & $0.070(79.5)$ & $0.013(61.6)$ & $0.012(57.3)$ & $0.055(35.3)$ & $0.040(79.7)$ & $0.009(61.5)$ & $0.007(57.2)$\\
$1.0$ & $0.097(37.0)$ & $0.074(84.0)$ & $0.015(67.1)$ & $0.013(63.2)$ & $0.063(40.1)$ & $0.043(84.1)$ & $0.010(67.0)$ & $0.008(63.1)$\\
\hline
\end{tabular}
%%%
\caption{\small NLO QCD+EW cross sections using the combination of Cut 3 and $|\Delta y_{Z,e}| < \Delta y_\text{cut}$ for 
various values of $\Delta y_\text{cut}$, separately for the $W^+ Z$ (left) and $W^- Z$ (right) processes. The acceptance $A$, 
the percentage of the cross section after applying the $\Delta y_{Z,e}$ cut, is provided in the parentheses.}
\label{tab:xs_Dy_Ze_Wpm}
\end{bigcenter}
\end{table}
%%%
%=========================================================================
\section{Conclusions}
\label{sect:conclusion}
We have presented new results for doubly-polarized cross sections of the $WZ$ production with fully 
leptonic decays at the LHC. Compared to the previous studies, two new kinematic cut setups are considered. 
These cuts are designed to observe the Radiation Amplitude Zero effect and to enhance the LL polarization. 
Our results show that the new cuts suppress the mixed polarizations drastically. 
For Cut 3, where $p_{T,WZ} < 70$~GeV and $p_{T,Z} > 200$~GeV, the dominant polarization is the TT 
with $f_{TT} \approx 66\%$ and the second one being the LL with $f_{LL} \approx 22\%$.
We found that one can suppress the TT to make the LL dominant by imposing a new 
cut of $|\Delta y_{Z,e}| < 0.5$. The nice feature of this cut is that the cross sections of all 
polarizations are large in the selected phase-space region, therefore we still have enough number of events 
for the analysis.  
%=========================================================================
\acknowledgments

We thank our experimental colleagues, in particular Joany Manjarr{\'e}s and Junjie Zhu, for useful discussions.
This research is funded by the Vietnam National Foundation for Science
and Technology Development (NAFOSTED) under grant number
103.01-2020.17.

%%%%%%%%%%%%
\clearpage
%The bibliography or References
\bibliographystyle{JHEP}
%\bibliography{main}

\providecommand{\href}[2]{#2}\begingroup\raggedright\endgroup
\end{document}